\documentclass{Interspeech2024}
\usepackage{tipa}
\usepackage[symbol]{footmisc}

\interspeechcameraready

\title{Towards EMG-to-Speech with a Necklace Form Factor}

\name[affiliation={1}]{Peter}{Wu$^*$}
\name[affiliation={1}]{Ryan}{Kaveh$^*$}
\name[affiliation={1}]{Raghav}{Nautiyal}
\name[affiliation={1}]{Christine}{Zhang}
\name[affiliation={1}]{Albert}{Guo}
\name[affiliation={1}]{Anvitha}{Kachinthaya}
\name[affiliation={1}]{Tavish}{Mishra}
\name[affiliation={1}]{Bohan}{Yu}
\name[affiliation={1}]{Alan W}{Black}
\name[affiliation={1}]{Rikky}{Muller$^{**}$}
\name[affiliation={1}]{Gopala Krishna}{Anumanchipalli$^{**}$}

\address{
  $^1$University of California, Berkeley}
\email{peterw1@berkeley.edu, ryankaveh@berkeley.edu}


\begin{document}

\maketitle
\def\thefootnote{*}\footnotetext{These authors contributed equally to this work. $^{**}$Equal advising.}\def\thefootnote{\arabic{footnote}}

\begin{abstract}
    
    Electrodes for decoding speech from electromyography (EMG) are typically placed on the face, requiring adhesives that are inconvenient and skin-irritating if used regularly. We explore a different device form factor, where dry electrodes are placed around the neck instead. 11-word, multi-speaker voiced EMG classifiers trained on data recorded with this device achieve 92.7\% accuracy. Ablation studies reveal the importance of having more than two electrodes on the neck, and phonological analyses reveal similar classification confusions between neck-only and neck-and-face form factors. Finally, speech-EMG correlation experiments demonstrate a linear relationship between many EMG spectrogram frequency bins and self-supervised speech representation dimensions. 
\end{abstract}

\section{Introduction}
Devices that decode speech from electromyography (EMG) are valuable for assistive communication applications, as they allow users to speak without vocalizing and can be life-changing in helping overcome dysarthria, dysphagia, stutters, and laryngectomies \cite{gaddy2021emg, Galego2021, Stepp2009, Wang2023, chan2001myo, meltzner2017silent, jou2006articulatory, lian2023unconstrained}. Prior work has synthesized high-fidelity speech from EMG, illustrating the viability of this technology for augmented spoken communication \cite{gaddy2021emg, Lyu2014, Wu2022}. Current devices measure articulatory signals from electrodes placed on the face \cite{gaddy2021emg, scheck2023emg_sil, kapur2018alterego}. We explore a more discreet form factor that only places dry electrodes around a neckband. This simplifies system setup, improves subject comfort, and eliminates the use of stigmatizing wet electrode arrays around the ears or face.

Surface EMG devices placed on the neck have shown success in diagnosing medical conditions like dysphagia \cite{vaiman2009surface, gupta1996surface}, estimating neck muscle activity \cite{sommerich2000use}, and measuring vocal fold vibrations \cite{lecluse1975electroglottography}. These techniques have been extended to speech classification using the same neck EMG signals, but high-accuracy speech decoding with wearables remains challenging \cite{chen2020content}. Additionally, existing demonstrations of surface EMG typically rely on wet electrodes that require periodic skin preparation to maintain consistent electrode-skin contact. While these wet electrodes can achieve consistent data with minimal electrode-motion-related artifacts, they are cumbersome and limit everyday use cases in public. Significantly more comfortable and easy-to-use 'dry' electrode devices have been demonstrated for arm-based EMG devices \cite{Laferriere2011, Myers2015}, but not yet on the neck. While these dry electrodes demonstrate higher electrode-skin impedances and are more susceptible to environmental interference and motion artifacts \cite{Chi2010}, their improved ease-of-use, lack of hydrogel, and increased longevity make them ideal for wearable speech decoding.

To explore wearable speech decoders, we propose a form factor with reusable dry electrodes and greater spatial coverage compared to traditional EMG neck devices, such as those used for electroglottographs \cite{lecluse1975electroglottography}. To this effect, this work presents a wireless, dry-electrode neckband used to capture EMG and speech data from two subjects. The resulting data is used for speech classification and mapping EMG to text and speech. Our high multi-speaker classification performance (Section \ref{sec:emg_classifier}) demonstrates an average accuracy of up to 92.7\% with just neck electrodes, comparable to numbers reported in prior work \cite{kapur2018alterego}. We also study the importance of the number of neck electrodes, finding that restricting this to two electrodes, as in electroglottographs \cite{lecluse1975electroglottography}, noticeably reduces classification accuracy. Phonological confusion analyses in Section \ref{sec:conf} indicates that classification behavior with neck-only data is similar to that with both neck and face data. Finally, speech-EMG correlation experiments in Section \ref{sec:speech_emg} indicate that our EMG data can be linearly mapped to speech acoustic representations and that neck electrodes can potentially be used on their own to perform speech decoding.

\section{Wireless EMG Neckband}

\begin{figure*}[t]
  \centerline{\includegraphics[width=6.5in]{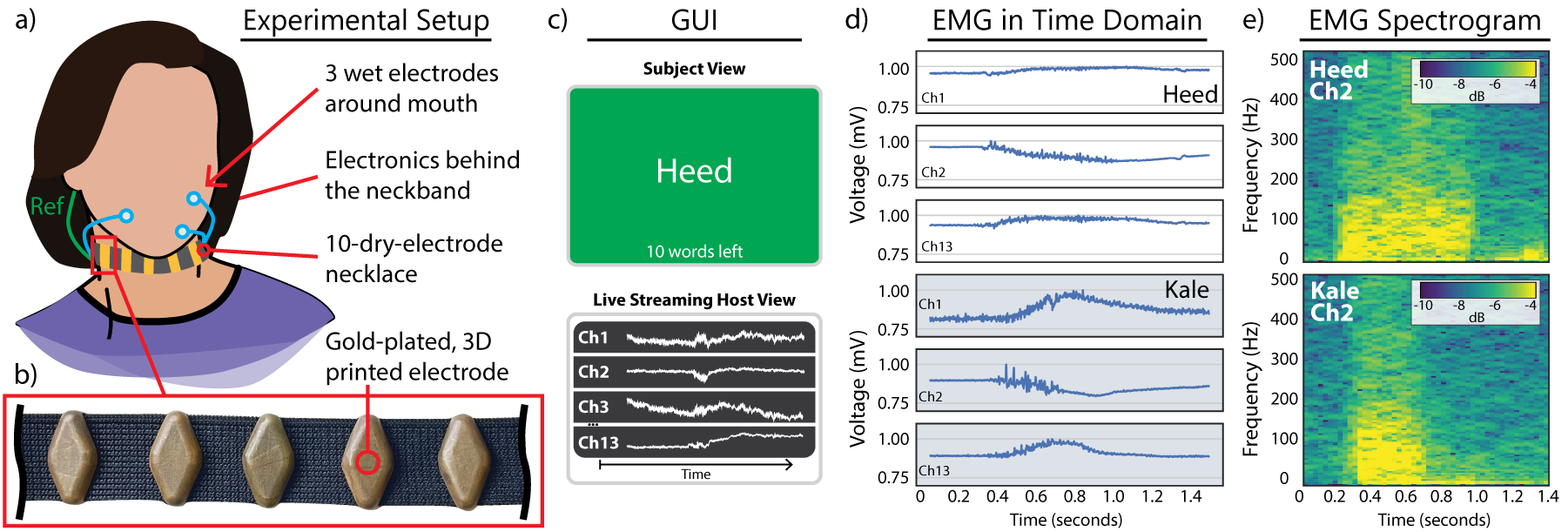}}
  \caption{(a) Experimental setup including a dry electrode neckband, baseline monitoring face electrodes, wet reference electrode behind the right ear, and neckworn electronics behind the head. (b) Partial photograph of 3D printed, gold plated neck electrodes. (c) Sample renders of the experiment GUI's subject and host views. Subject view displays a teleprompter while raw EMG data is live plotted on the host view. (d) Raw sample EMG from a single utterance of the words 'Heed' and 'Kale'.  (d) Sample EMG time-frequency spectrograms (see section 3.2) from a single utterance of the words 'Heed' and 'Kale'.}
  \label{fig:system_setup}
  \vspace{-15pt}
\end{figure*}

The wireless recording setup was designed to explore easy-to-use, discreet, and comfortable wearables for speech decoding. As a result, the system consists of a dry electrode EMG neckband and a wireless recording module (Fig. \ref{fig:system_setup}a and b). In addition, three wet Ag/AgCl electrodes are included on the face to simultaneously record a benchmark for the neck recorded EMG. Emphasis was placed on modularity, and system characteristics such as electrode count, location, and chemistry can be easily augmented due to our 3D printing-based fabrication process and recording hardware that can record electrophysiological signals from up to 64 electrodes.

\subsection{Electrodes}

The dry electrode necklace comprises 10 equally spaced dry electrodes placed all around an elastic neckband (Fig. \ref{fig:system_setup}b). The dry electrodes are 3D printed and electroless gold-plated to achieve an inert, biocompatible surface \cite{Kaveh2022}. This fabrication process results in dry electrodes with a high effective surface area that can be used without any hydrogel application or skin abrasion, greatly improving user comfort while also reducing the risk of skin irritation \cite{Stjerna_2010}. Lastly, this fabrication process can easily be repeated or augmented to result in arbitrarily shaped electrodes that conform to the jaw or other features in future studies.

The electrodes were arranged to comfortably fit the average neck circumference of men (40.1 cm +/- 3.05 cm) and women (34.8 cm +/- 2.79 cm) \cite{Kornej2022}. As a result, the final neck electrodes have a cross-sectional surface area of 2 $cm^2$ and a pitch of 3 $cm$. These electrodes achieve an average 50 Hz electrode-skin impedance (ESI) and phase of 55.1 $k\Omega s$ and $-95^ \circ$ at 50 Hz (N = 10) which, while potentially higher than the ESI of wet electrodes, is well within the input parameters of the system's neural recording frontends. The electrodes exhibit a mean electrode DC offset of -13.3 $mV$ with a standard deviation of 14.1 $mV$ (N = 600). After fabrication, the electrodes are clipped into an elastic, velcro neckband. Each electrode is soldered to a 36 AWG jumper cable that is threaded through the elastic band to minimize wire-motion-related artifacts.

The neck electrodes are used in conjunction with three wet, Ag/AgCl electrodes around the subject's lips to track lip, palate, and jaw movements \cite{gaddy2021emg}. Two electrodes are placed above and below the left side of the subject's lips, while one electrode is placed on the right side of the lips. These wet electrodes are used to provide a comparison benchmark for decoding tasks described in section \ref{sec:speech_repr} and \ref{sec:results}. All neck and face electrodes are used to perform differential measurements with the shared reference wet electrode placed on the subject's right mastoid. Each differential channel is connected to a wireless recording module worn behind the neck.

\subsection{Wireless Recording Module}

EMG was recorded using an existing compact recording platform, known as the miniature, wireless, artifact-free neuromodulation device (WANDmini) \cite{Zhou2019}. WANDmini, originally built for implanted neural recording, has been adapted for wearable applications and deployed in multiple electrophysiological studies \cite{Kaveh2020, Moin2021}. Due to being small (2.5~$\times$~2.5~cm\textsuperscript{2}) and lightweight (3.8 g), WANDmini can discreetly fit on the back of the neckband and be comfortably worn for hours. WANDmini records and digitizes EMG signals with a custom neuromodulation IC \cite{Johnson2017} (NMIC, Cortera Neurotechnologies, Inc.). The recorded data is then processed and packetized by an onboard FPGA SoC (166~MHz ARM Cortex M3 processor - SmartFusion2 M2S060T, Microsemi). The packetized data is then transmitted to the base station by a 2.4~GHz BLE radio (nRF51822, Nordic Semiconductor) that is also used to configure WANDmini. When powered by a 3.7~V, 300~mAh lithium polymer (LiPo) battery, the neckband and WANDmini can operate and digitize 64 channels of data for roughly 24~hours. Pertinent system specifications are described in Table \ref{tb:wand}.

Packetized EMG signals are received by a wireless base station connected to a laptop running a Python graphical user interface (GUI) that not only provides real-time data visualization but also provides cues and teleprompting for subjects (Fig. \ref{fig:system_setup}c).
In addition to data visualization and teleprompting, the GUI also records audio of the subject vocalizing the cued utterance. All audio and EMG are synchronized with trigger signals sent by the GUI to the laptop's microphone and WANDmini. During each experiment, the GUI consists of two main components: a screen displaying real-time EMG signals from the EMG device, and a teleprompter that presents words sequentially from a given utterance list. This setup allows the subject to vocalize the words shown while a host monitors the recorded EMG and audio data. The teleprompter’s pacing can be adjusted to accommodate different speakers, ensuring that the speech produced during data collection is natural.

\begin{table}[t]
\setlength{\tabcolsep}{17pt}
\caption{WANDmini System Electrical Specifications}
\vspace{-20pt}
\begin{center}
\begin{tabular}{rc}
\toprule 
Maximum Recording Channels      & 64       \\
Recording Channels Used         & 13       \\
Reference Location              & Right Mastoid \\
Input Range                     & 100 mVpp \\
ADC Resolution                  & 15 bits  \\
ADC Sample Rate                 & 1 kSps   \\
Noise Floor                     & $70 nV/\sqrt{Hz}$ \\
Wireless Data Rate              & 2 Mbps    \\ 
WANDmini Power                  & 46 mW     \\
Weight (w/o battery)            & 3.8 g                  \\ 
Battery Life                    & ~44 Hours \\\bottomrule\hline
\end{tabular}
\label{tb:wand}
\end{center}
\vspace{-25pt}
\end{table}

\subsection{Experimental Setup and Data Collection}
\label{sec:data_collection}

To verify the neckband system performance, electrophysiological and speech measurements were performed on two subjects. At the start of each experiment, the subject would arrive and don the neckband. The two center electrodes of the neckband were aligned around the subject's Adam's apple and then tightened so each electrode was making contact with the skin. After the neck electrodes were placed around the entire neck, the wet Ag/AgCl electrodes were placed around the subject's lips. Once all the electrodes were connected to the recording frontend, WANDmini would be stowed in a 3D-printed enclosure affixed to the back of the neckband. The subject would then sit in front of the laptop, approximately 3 feet away from the microphone, and vocalize the utterances displayed on the screen (\ref{fig:system_setup}c). In total, 11 words were displayed 10 times each (each instance of a word is referred to as an utterance). The GUI first shows 'wait' for three seconds, then the specified word for three seconds, and a final 'wait' for three seconds. As a result, each recording totals 9 seconds but is sliced to the 1.5 seconds around the actual utterance. This user study was approved by UC Berkeley’s Institutional Review Board (CPHS protocol ID: 2018-09-11395).

\section{Computational Methods}

\subsection{Dataset}

Our complete dataset is composed of 13 channels of EMG (sampled at 1000 Hz) and voiced acoustics (sampled at 44100 Hz) for 11 words. These 13 channels comprise 10 neck channels and 3 face channels. This dataset was further segmented into a 10-channel dataset (consisting only of the dry neck electrodes), and a 13-channel dataset (consisting of all electrodes). 

Our utterance list consists of the following words, listed with their IPA transcriptions: heed \textipa{[hid]}, had \textipa{[h\ae d]}, hood \textipa{[hUd]}, tail \textipa{[t\super{h}eIl]}, kale \textipa{[k\super{h}eIl]}, doe \textipa{[doU]}, goat \textipa{[goUt]}, aba \textipa{[aba]}, ada \textipa{[ada]}, aga \textipa{[aga]}, and aka \textipa{[ak\super{h}a]}. In this set of utterances, we have minimal pairs that differ only in the vowel (\textipa{[hid]/[h\ae d]/[hUd]}), (near-)minimal pairs that differ only in plosive place of articulation (\textipa{[aba]/[ada]/[aga]} and \textipa{[doU]/[goUt]} and \textipa{[t\super{h}eIl]/[k\super{h}eIl]}), and minimal pairs that differ in voicing and aspiration of a plosive (\textipa{[aga]/[ak\super{h}a]}). Minimal pairs in our data allow us to isolate certain aspects of speech while keeping other factors constant, which helps us assess the performance of our model on each of these aspects.

Each word is recorded 10 times by two native male English speakers (arbitrarily fixed as Speaker 1 and Speaker 2). The pertinent 1.5 seconds recorded for each utterance are kept, yielding 5 minutes and 30 seconds of data in total. Since EMG and voiced speech audio are recorded at the same time (Section \ref{sec:data_collection}), we treat these two modalities as temporally aligned.

\subsection{EMG Representations}
\label{sec:emg_repr}

Raw EMG in the time-domain can be readily featurized for classification tasks. Inspired by openSMILE, we compute the following statistics for each EMG channel: max, min, range, max position, min position, arith-mean, quad-mean, std, var, kurtosis, skewness, 25-percentile, 75-percentile, number of peaks, mean peak amplitude, mean abs slope, rise time, fall time, zcr, mcr \cite{eyben2010opensmile}. We concatenate the vectors from each dimension into a single vector, giving us a feature vector for each EMG utterance. Sample raw EMG data of a single utterance of the words 'Heed' and 'Kale' are shown in Figure \ref{fig:system_setup}d.

While the above time-domain and statistical features are adequate for classification, they are also more noise susceptible and contain fewer dimensions relative to frequency-domain representations of EMG. A time-frequency spectrogram, on the other hand, can be easily filtered and readily compared to acoustic speech representations. As a result, spectrograms are computer for every utterance's channels using consecutive Fourier transforms. The number of samples per segment was set to 100, while the overlap between segments was set to 50. Additionally, the number of Fast Fourier Transform (FFT) points per segment was set to 128. These parameters allowed for a high quality spectrogram to be generated in order to better enable data analysis. Figure \ref{fig:system_setup}e depicts two spectrograms extracted from channel 2 of the raw EMG shown in Figure \ref{fig:system_setup}d.


\subsection{Self-Supervised Acoustic Representations}
\label{sec:speech_repr}

Self-supervised acoustic speech representations encode waveforms into a dense vector space suitable for speech production and perception \cite{Chen2021WavLMLS, yang21c_superb}. Given the versatility of mapping to and from these representations, we compare our EMG data with these features to study the feasibility of mapping EMG signals to acoustic ones. We choose WavLM as the self-supervised speech representation in this paper due to its success in speech benchmarks and articulatory tasks \cite{Chen2021WavLMLS, yang21c_superb, cho2023evidence, wu2023speechema}. WavLM is a state-of-the-art pre-trained Transformer model that extends HuBERT, a masked speech prediction method, by adding a denoising objective \cite{vaswani2017transformer, hsu2021hubert}. This model also employs a gated relative position bias to better utilize sequence ordering and is trained on a larger and more diverse dataset across different scenarios and languages \cite{Chen2021WavLMLS}. Given the generalizability of WavLM, we hypothesize that features extracted by this model may be robust enough to map to EMG data. Since WavLM accepts 16000 Hz waveform inputs, we downsample our speech audio from 44100 Hz. We extract features from the tenth layer of WavLM, given the articulatory properties around that model depth \cite{cho2023evidence}. This yields 1024-dimensional vectors at each time step, where time steps have a sampling rate of 50 Hz.

\begin{figure}[t]
  \centerline{\includegraphics[width=3in]{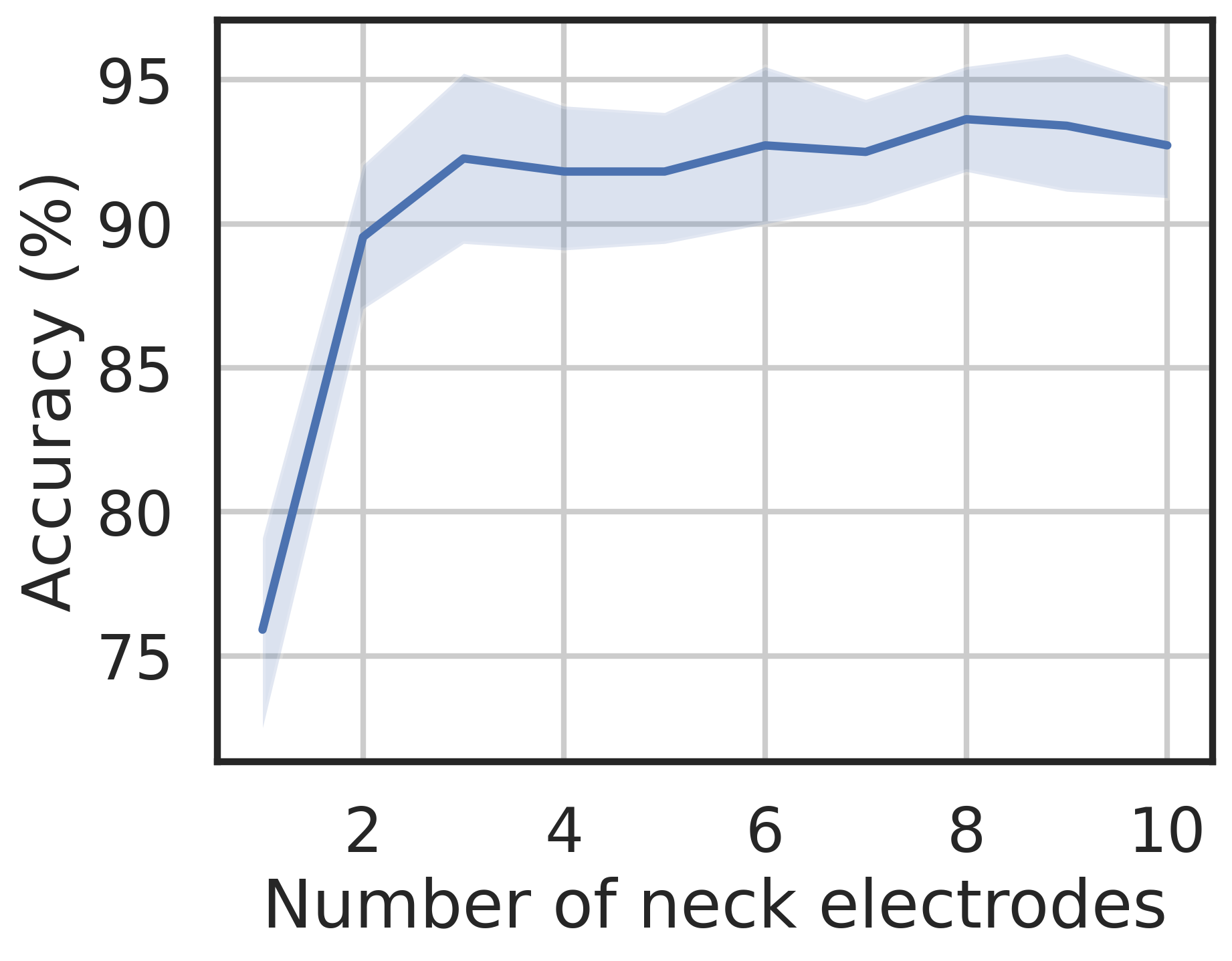}}
  \caption{Classification accuracy for different numbers of neck electrodes. Solid lines are means and opaque regions are 95\% confidence intervals.}
  \label{fig:one_shot}
  \vspace{-15pt}
\end{figure}

\section{Results}
\label{sec:results}

\subsection{EMG Classification}
\label{sec:emg_classifier}

To check our EMG data quality, we first classify words from EMG with a random forest classifier with max depth of 32. Here, our inputs are the EMG statistics vectors described in Section \ref{sec:emg_repr}. On 10 random 80\%-20\% train-test splits of our two-speaker dataset, we achieve a mean accuracy of 93.9\% with a 95\% confidence interval of [92.7\%, 95.0\%] using all 13 electrodes. With only the 10 neck electrodes, we achieve a mean accuracy of 92.7\% with a 95\% confidence interval of [90.9\%, 94.8\%]. Accuracies in both cases are much higher than chance, which is around 9\% given that each of our 11 words has the same number of utterances. This suggests that our EMG data and the neck-only subset both contain useful linguistic content. As a comparison with prior work, classification with only face electrodes here yields a mean accuracy of 88.6\% with a 95\% confidence interval of [85.5\%, 91.8\%], compared to 92\% in prior work \cite{kapur2018alterego}.

We also check the importance of the number of neck electrodes for EMG decoding. With the aforementioned train-test splits, we calculate the word classification accuracy given different numbers of neck electrodes as input. Accuracy means and 95\% confidence intervals are plotted on Figure \ref{fig:one_shot}. Accuracy noticeably improves after adding the second and third electrode, and continues improving a bit up to adding the eighth electrode. This suggests that language decoding from EMG can improve when the device has more than two electrodes, such as in the typical electroglottograph setup\cite{lecluse1975electroglottography}.

\begin{figure}[tbp] 
  \centerline{\includegraphics[width=3.5in]{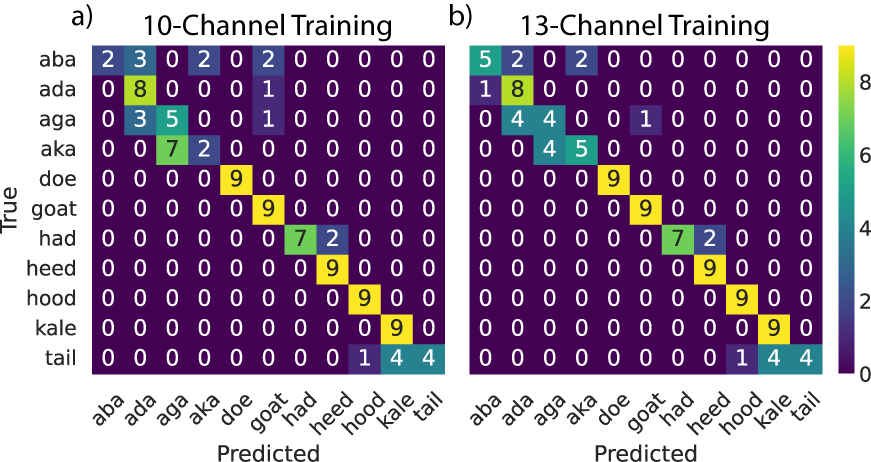}}
  \caption{Confusion matrices using model trained on (a) the 10 neck channels and (b) all 13 channels.}
  \label{fig:confusions}
  \vspace{-15pt}
\end{figure}

\subsection{Phonological Confusion}
\label{sec:conf}

To analyze our EMG data phonologically, we generate confusion matrices for EMG classification (Fig. \ref{fig:confusions}). We use the same random forest classifier, utterance set, statistical EMG input vector representations, and target labels as in Sec. \ref{sec:emg_classifier}. In order to emphasize model confusion, we report results for 1-shot classification. Specifically, we train on all of one speaker's data and 1 random utterance for each word from the other speaker, and test on the remaining data from the second speaker. We generate two confusion matrices, one for each first-second speaker assignments, and add together the matrices element-wise for conciseness. This process was performed with the full 13-channel dataset (Fig. \ref{fig:confusions}a) and with the 10-channel dataset (Fig. \ref{fig:confusions}b).

As is can be seen from the confusion matrices in Figure \ref{fig:confusions}, our model can generally predict the correct vowel based on the EMG data, but it has more trouble with identifying specific consonants. In particular, we observe that the model trained on all 13 channels primarily confuses plosives that either: (1) differ in place of articulation but match in voicing i.e. \textipa{[d]}/\textipa{[b]} and \textipa{[k]}/\textipa{[t]}, or (2) differ in voicing but match in place of articulation i.e. \textipa{[k]}/\textipa{[g]}. In addition, the model may also be weaker at differentiating between front vowels as opposed to other vowels as the model confuses had and heed but can distinguish between these words and hood. This may be because of the placement of electrodes on the neck given that back vowels cause more muscles to be engaged near the neck while front vowels do not.

We note that there is some improvement in model performance after adding three extra wet electrodes to the face. The model trained on only neck electrode channels confuses \textipa{[goUt]} and \textipa{[aba]} while the model trained on all 13 channels does not. The 3 extra electrodes near the lips may help the model identify the labial consonant \textipa{[b]} in \textipa{[aba]}. In addition, for the ground truth word \textipa{[ada]}, the 13-channel model predicts \textipa{[aba]} while the 10-channel model predicts \textipa{[goUt]}. The facial electrodes likely also help the model distinguish between vowels and consonants since the 13-channel model is able to identify a word-initial vowel as opposed to the 10-channel model, which confuses them. Generally, the model confusability is similar for the 13- and 10-channel settings, suggesting that our necklace form factor may capture enough information to decode speech.

\subsection{Speech-EMG Correlation}
\label{sec:speech_emg}

\begin{figure}[tbp]
  \centerline{\includegraphics[width=2.7in]{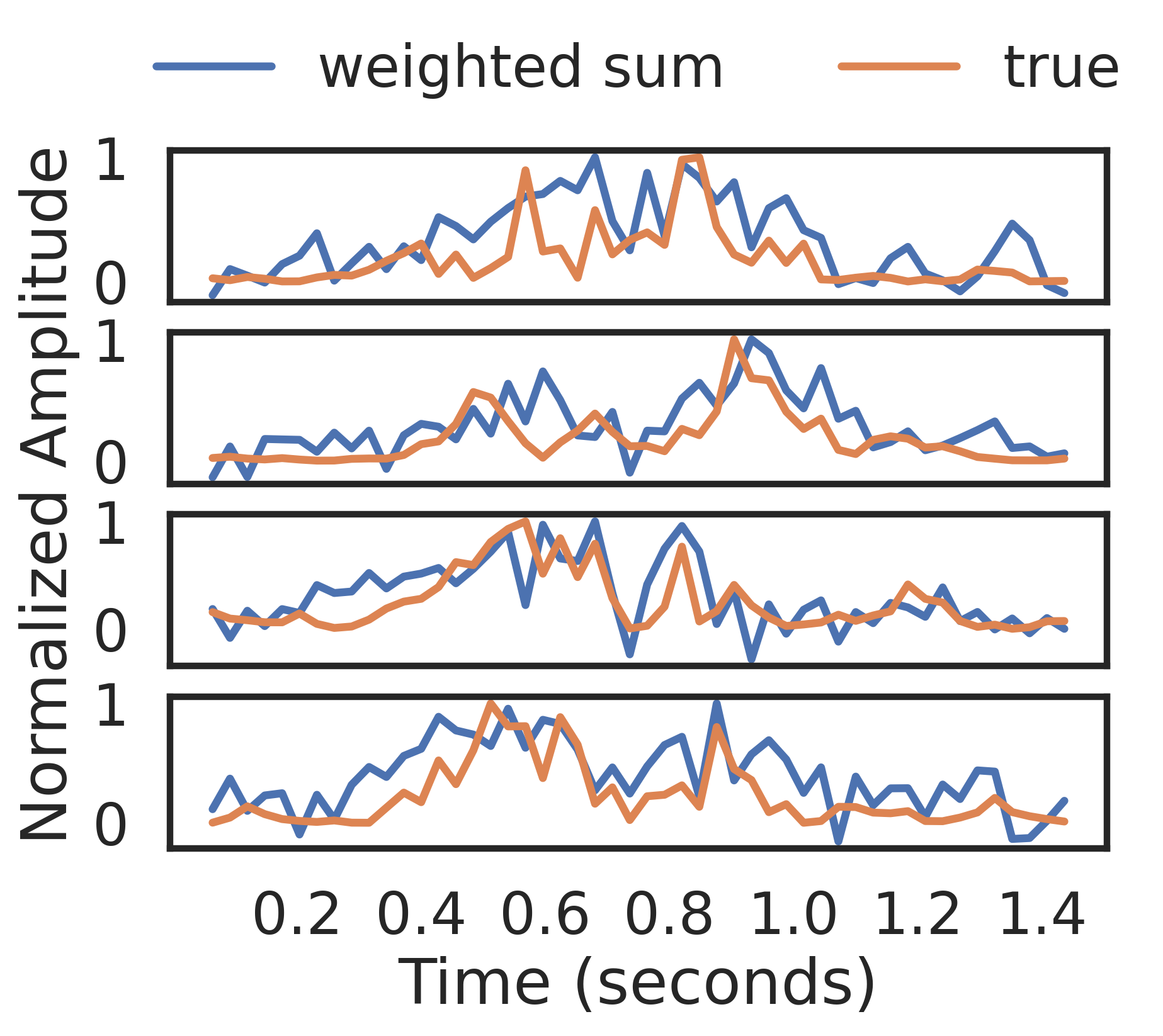}}
  \caption{Weighted sum of self-supervised speech features match EMG spectrogram frequency bins. Here, we plot 1 EMG channel of a "kale" utterance for bins 90-94 Hz, 102-105 Hz, 238-242 Hz, and 348-352 Hz (Top-to-bottom).}
  \label{fig:speech_emg}
  \vspace{-15pt}
\end{figure}

Through a speech-EMG correlation experiment, we observe that self-supervised speech features and time-frequency representations of EMG correlate noticeably. For all utterances in our dataset, we encode waveforms into WavLM representations \cite{Chen2021WavLMLS} (Section \ref{sec:speech_repr}) and EMG into spectrograms (Section \ref{sec:emg_repr}). For our EMG representation, we flatten the 10 EMG channels and 129 spectrogram frequencies into a 1290 vector, yielding 1290 values varying over time. We linearly interpolate the WavLM features to match the length of the EMG features. Then, we train a linear regression model on all of the data to map the WavLM vector at each time step to the respective flattened EMG vector. In other words, we approximate each flattened EMG dimension with a weighted sum of WavLM dimensions. Out of the 1290 EMG dimensions, 33.1\% of them have a mean Pearson correlation coefficient of at least 0.5 with a linear combination of WavLM dimensions, with examples visualized in Figure \ref{fig:speech_emg}. 33.1\% is noticeably higher than the 0.0\% that occurs when we replace WavLM elements with numbers randomly uniformly sampled from [0,1]. This suggests that our EMG data contains useful speech acoustic information.

\section{Conclusion}

This work presents a discreet EMG neckband with reusable dry electrodes capable of performing speech classification and analysis. Ablation studies with our device indicate that the neck electrodes can achieve a high classification accuracy on their own (92.7\%) which is similar to classification accuracies achieved with both neck and face electrodes (93.9\%). Additionally, speech-EMG correlation experiments reveal that our device can record useful speech information for further speech decoding work. Moving forward we will collect sentence-length utterances from a larger set of speakers to explore wearable EMG-to-speech synthesis through the use of a necklace form factor without any face electrodes.



\bibliographystyle{IEEEtran}
\bibliography{mybib}

\end{document}